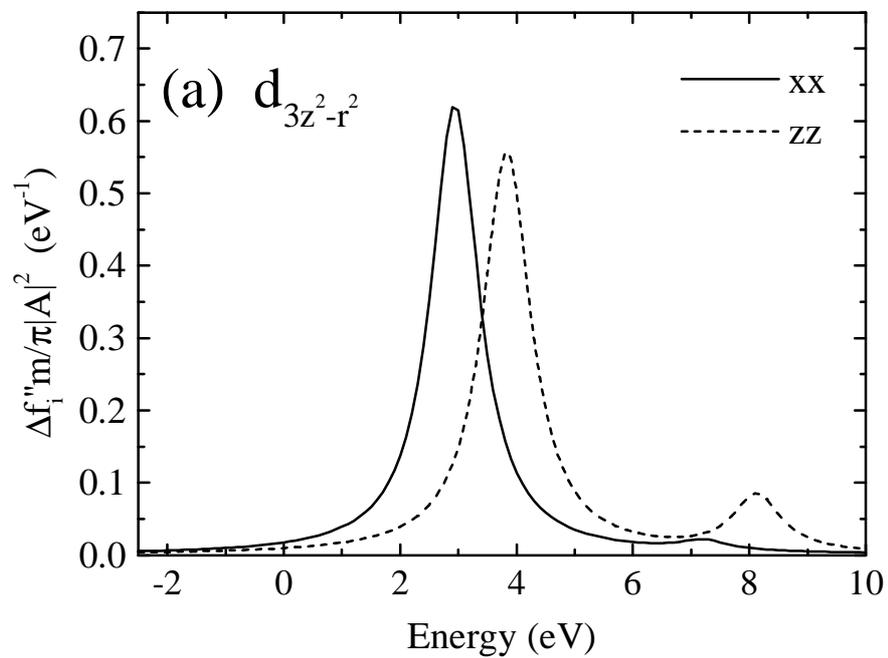

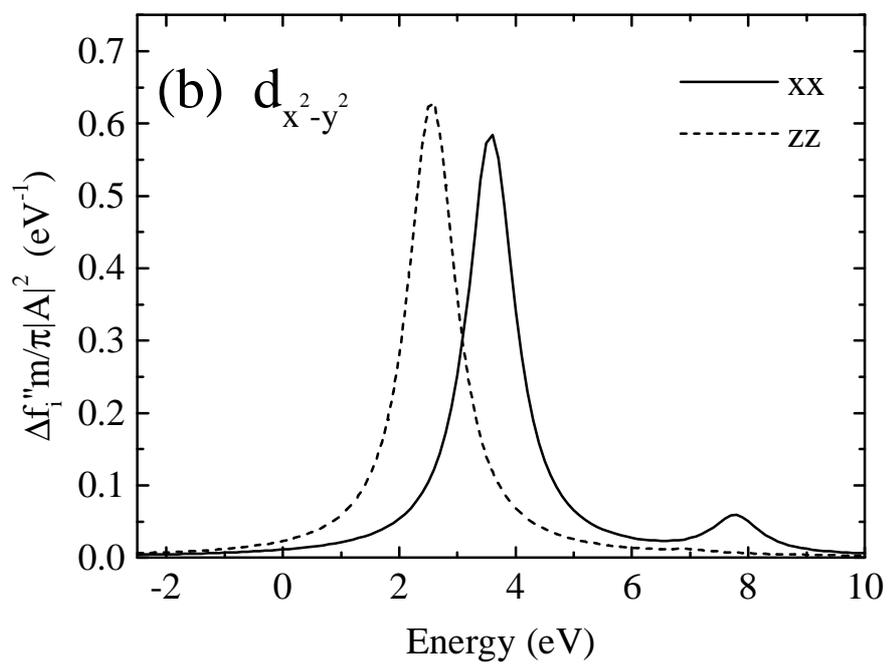

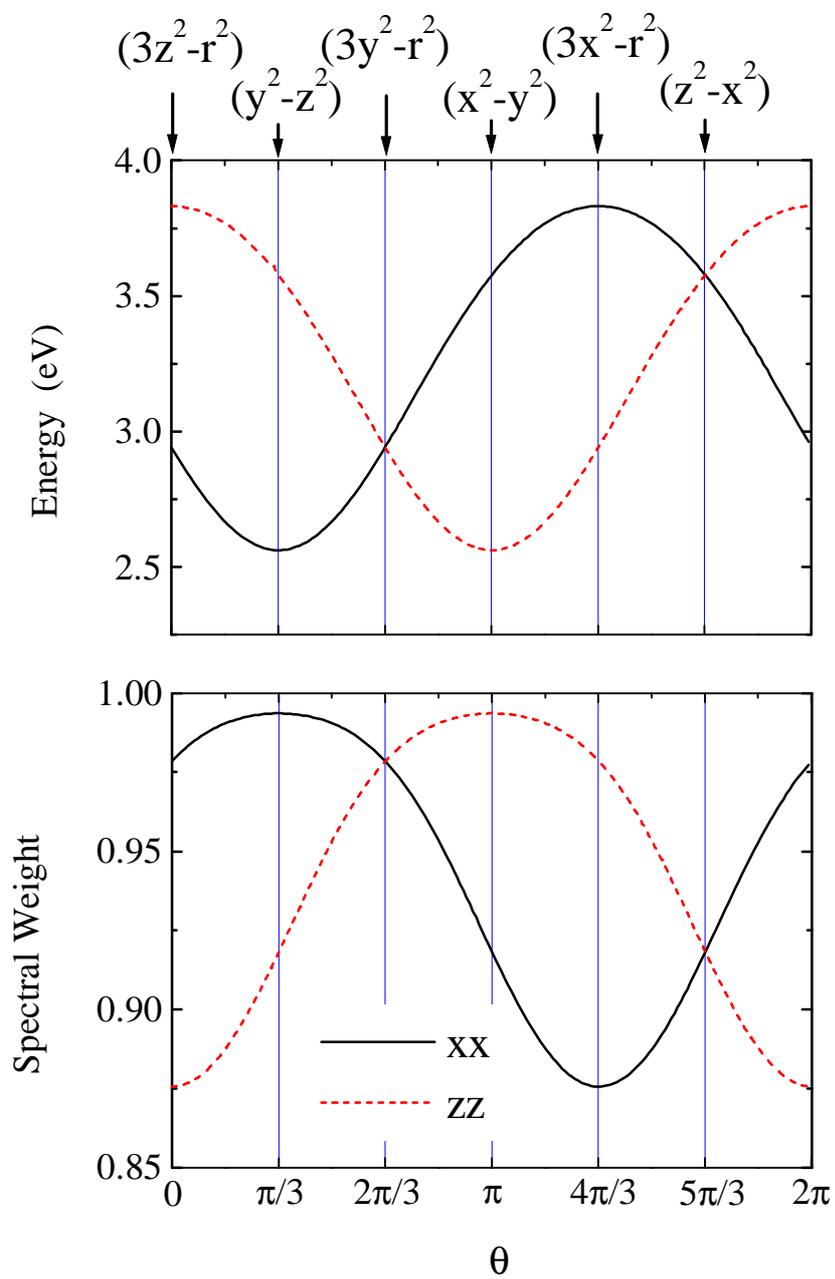

# Theory of Anomalous X-ray Scattering in Orbital Ordered Manganites

Sumio Ishihara and Sadamichi Maekawa
*Institute for Materials Research, Tohoku University, Sendai, 980-77 Japan*
(December 15, 1997)

We study theoretically the anomalous X-ray scattering as a new probe to observe the orbital orderings and excitations in perovskite manganites. The scattering matrix is given by the virtual electronic excitations from Mn $1s$ level to unoccupied Mn $4p$ level. We find that orbital dependence of the Coulomb interaction between Mn $3d$ and Mn $4p$ electrons is essential to bring about the anisotropy of the scattering factor near the K edge. The calculated results in $MnO_6$ clusters explain the forbidden reflections observed in $La_{0.5}Sr_{1.5}MnO_4$ and $LaMnO_3$. A possibility of the observation of the orbital waves by the X-ray scattering is discussed.

71.10.-w, 75.90.+w, 78.70.Ck

The discovery of high $T_c$ cuprates has stimulated intensive study of transition-metal oxides (TMO) from the new theoretical and experimental view points. One of the key factors to study the electronic structures in TMO is the orbital degrees of freedom in $3d$ transition-metal ions. In particular, the orbital degrees in perovskite manganites and related compounds bring about a lot of fruitful and dramatic phenomena [1–5]. The electron configuration in a $Mn^{3+}$ ion is $(e_g)^1(t_{2g})^3$ with parallel spins due to the strong Hund coupling. In the cubic oxygen octahedral cage, the two $e_g$ levels are degenerate. Thus, an $e_g$ electron has the orbital degrees of freedom as well as charge and spin. In order to reveal the unique magneto-transport phenomena in manganites, it is essential to study the nature of the orbital degrees and their correlation with spin and lattice. However, the experimental techniques directly detecting the orbital ordering have been limited [6].

Recently, Murakami et al. have applied the anomalous X-ray scattering in order to observe the orbital ordering in the single layered manganites $La_{0.5}Sr_{1.5}MnO_4$ [7]. By utilizing the anisotropy of the tensor of susceptibility (ATS) reflection [8,9], they observed a sharp $(3/4,3/4,0)$ reflection, which is the forbidden one, at the $Mn^{3+}$ K edge. This experimental result implies the followings: (i) There exist two non-equivalent $Mn^{3+}$ ions in the $MnO_2$ plane, which exhibit an alternating orbital ordering. The ordering is called as antiferro-type hereafter. (ii) The electric dipole (E1) transition from Mn $1s$ core level to unoccupied Mn $4p$ level causes the anomalous scattering. This method was also applied to the undoped manganite $LaMnO_3$ [10] where the $(3,0,0)$ forbidden reflection suggesting the antiferro-type orbital ordering was observed. The new experimental technique not only confirms the orbital ordering in the two manganites, but has a large potential for applications to study the nature of the orbital in a variety of TMO.

In this letter, we study theoretically the anomalous X-ray scattering as a new probe to detect the orbital ordering in manganites. We identify the origin of the anisotropy of the scattering factor in the orbital ordered state. The orbital dependence of the Coulomb interactions between $3d$ and $4p$ electrons is dominant on ATS. We also discuss a possibility to detect the orbital waves by the inelastic X-ray scattering.

First, we briefly mention the general formula of the X-ray scattering factor. In the X-ray scattering process, the structure factor is defined by the scattering amplitude divided by $(e^2/mc^2)$. By the second-order perturbational calculation with respect to the electron-photon interaction, the anomalous part of the scattering factor for the $i$-th ion in the unit cell is given by [11,12],

$$\Delta f_i(k', k'')_{\alpha\beta} = \frac{m}{e^2} \sum_l \left\{ \frac{\langle f|j_{i\alpha}(-\vec{k}')|l\rangle\langle l|j_{i\beta}(\vec{k}'')|0\rangle}{\varepsilon_0 - \varepsilon_l - \omega_{k''} - i\delta} + \frac{\langle f|j_{i\beta}(\vec{k}'')|l\rangle\langle l|j_{i\alpha}(-\vec{k}')|0\rangle}{\varepsilon_0 - \varepsilon_l + \omega_{k'} - i\delta} \right\}, \quad (1)$$

where the electronic system is excited from the initial state $|0\rangle$ with energy $\varepsilon_0$ to the intermediate one $|l\rangle$ with $\varepsilon_l$, and is finally relaxed to the final state $|f\rangle$ with $\varepsilon_f$. Here, $\alpha$ and $\beta$ represent the polarization of photons, $\omega_{k'(k'')}$ is the incident (scattered) photon energy with momentum $k'(k'')$ and $\delta$ is an infinitesimaly small constant. The current operator describing the $1s \rightarrow 4p$ E1 transition is given by $j_{i\alpha}(\vec{k}) = \frac{eA}{m}\sum_\sigma P^\dagger_{i\alpha\sigma}s_{i\sigma}$ where $P^\dagger_{i\alpha\sigma}$ and $s_{i\sigma}$ are the creation operator of Mn $4p$ electron and the annihilation of Mn $1s$ core one, respectively, with spin $\sigma$ and orbital $\alpha$. $A$ is the coupling constant given by $A = \int d\vec{r} e^{i\vec{k}\vec{r}} \phi_{4p\alpha}(\vec{r})^*(-i\nabla_\alpha)\phi_{1s}(\vec{r})$ where $\phi_m(\vec{r})$ ($m = 1s, 4p$) is the atomic wave function. The quadrupole transition is ruled out in Eq.(1), since it is expected to be weak near the K edge in manganites [13,14].

Concerning the anomalous part of the structure factor, it is worth to note the followings: (i) By utilizing the fact that $\Delta f_i$ is the second rank tensor with respect to photon polarization, we can directly detect the anisotropy of the microscopic electronic structure [8], i.e., the microscopic birefringence and dichroism due to the orbital order. (ii) The imaginary part of $\Delta f_i$, $\Delta f_i''$, with $|0\rangle = |f\rangle$ is written as $(\pi m/e^2)\sum_l \delta(\varepsilon_0 - \varepsilon_l + \omega_k)|\langle 0|\sum_\sigma j_{i\alpha}(\vec{k})|l\rangle|^2$ near the K edge. For the uniform ordering of orbitals which we call

ferro-type, this is proportional to the intensity of the Mn $1s$ X-ray absorption spectra (XAS) or X-ray fluorescence spectra. On the other hand, for the antiferro-type orbital ordering, it is not possible to detect the anisotropy of $\Delta f_i$ by XAS, in contrast to the diffraction measurement of present interest.

As mentioned above, the anomalous scattering is dominated by the $1s \rightarrow 4p$ E1 transition. In this case, how does the $3d$ orbital orderings reflect on the anisotropy of the anomalous scattering factor? In order to study the problem, let us consider the electronic structure in the MnO$_6$ octahedron, since the local electronic excitation dominates $\Delta f_i$. In the Mn ion, a minimal set of the orbital is $\{1s, 3d_\gamma \ (\gamma = \gamma_{\theta+}, \gamma_{\theta-}), 4p_\gamma \ (\gamma = x, y, z)\}$, where $|3d_{\gamma_{\theta+}}\rangle = \cos(\theta/2)|3z^2 - r^2\rangle + \sin(\theta/2)|x^2 - y^2\rangle$ and $|3d_{\gamma_{\theta-}}\rangle$ is its counterpart. Six O $2p$ orbitals, which constitute to the $\sigma$-bond with the Mn orbitals, are recombined by the symmetry in $O_h$ group as $\{2p_{\gamma_{\theta+}}, 2p_{\gamma_{\theta-}}, 2p_x, 2p_y, 2p_z, 2p_{r^2}\}$, where $x$ etc. represent the bases of the irreducible representation in the group. Then, it is concluded that the electron hybridizations between Mn $3d$ and O $2p$ orbitals and between the Mn $4p$ and O $2p$ orbitals do not result in ATS.

One of the promising origin of ATS is the Coulomb interactions between Mn $3d$ and $4p$ electrons. When we consider that one of the $e_g$ orbitals $3d_{\gamma_{\theta+}}$ is occupied, the Hartree potential breaks the cubic symmetry and thus lifts the degeneracy of Mn $4p$ orbitals. The interaction between Mn $3d$ and $4p$ electrons is represented as

$$V(3d_{\gamma_{\theta+}}, 4p_\gamma) = F_0 + 4F_2 \cos\left(\theta + m_\gamma \frac{2\pi}{3}\right), \quad (2)$$

where $m_x = +1$, $m_y = -1$, and $m_z = 0$. $F_n$ is the Slater-integral between $3d$ and $4p$ electrons. The explicit formula of $F_n$ is given by $F_0 = F^{(0)}$ and $F_2 = \frac{1}{35}F^{(2)}$ with $F^{(n)} = \int dr dr' r^2 r'^2 R_{3d}(r)^2 R_{4p}(r')^2 \frac{r_<^n}{r_>^{n+1}}$ where $r_<$ ($r_>$) is smaller (larger) one between $r$ and $r'$. When $d_{3z^2-r^2}$ orbital is occupied ($\theta = 0$), the Hartree energy of $4p_z$ orbital is higher than that of $4p_{x(y)}$ orbital by $6F_2$. As a result, $(\Delta f_i)_{xx(yy)}$ dominates the anomalous scattering near the edge in comparison with $(\Delta f_i)_{zz}$.

The inter-atomic Coulomb interaction between Mn $4p_\gamma$ and O $2p_{\gamma_{\theta+}}$ electrons also provides an origin of ATS through the Mn $3d$ - O $2p$ hybridization. The interaction is represented by $V(2p_{\gamma_{\theta+}}, 4p_\gamma) = \varepsilon + \frac{\varepsilon \rho^2}{5} \cos\left(\theta + m_\gamma \frac{2\pi}{3}\right)$, where the definition of $m_\gamma$ is the same as that in Eq.(2). $\varepsilon = Ze^2/a$ and $\rho = \langle r_{4p}\rangle/a$ where $Z = 2$, $a$ is the Mn-O bond length, and $\langle r_{4p}\rangle$ is the average radius of Mn $4p$ orbital. Although the above two interactions cooperate to bring about ATS, it seems likely that magnitude of $V(2p_{\gamma_{\theta+}}, 4p_\gamma)$ is much reduced by the screening effects, in comparison with $V(3d_{\gamma_{\theta+}}, 4p_\gamma)$.

The lattice distortion in the oxygen octahedron also lifts the degeneracy of Mn $4p$ orbitals. The difference of the Hartree energy between Mn $4p_{x(y)}$ and Mn $4p_z$ orbitals is given by $V(4p_{x(y)}) - V(4p_z) = 18\varepsilon\rho^2/5(\delta a/a)$ with $\delta a = a_z - a_{x(y)}$, where $a_\alpha$ is the Mn-O bond length in the $\alpha$-direction. In La$_{1-x}$Sr$_{1+x}$CuO$_4$ and La$_{1-x}$Sr$_{1+x}$NiO$_4$ [15–18], this contribution seems to dominate the anisotropy in XAS. On the other hand, in La$_{1-x}$Sr$_{1+x}$MnO$_4$ at $x \sim 0.5$, it is confirmed that the notable lattice distortion in the MnO$_6$ octahedron is not observed experimentally [19,20]. Therefore, the mechanism based on Eq. (2) mainly provides the origin of ATS. In LaMnO$_3$ where $\delta a/a$ is about 13% [21], the contribution from the lattice distortion weakens the anisotropy caused by the intra- and inter-atomic Coulomb interactions.

In order to confirm the mechanism of ATS discussed above, we calculate the anisotropy of $\Delta f_i$ in a MnO$_6$ cluster. The following Hamiltonian is adopted in the calculation [22,24];

$$H = H_0 + H_t + H_{core} + H_{3d-4p} + H_{3d-3d}. \quad (3)$$

$H_0$ and $H_t$ describe the energy level and the electron transfer between Mn $3d_\gamma$ and O $2p_\gamma$ orbitals, respectively. The core hole potential is introduced in $H_{core}$. $H_{3d-4p}$ represents the Coulomb interaction between Mn $3d$ and $4p$ electrons shown in Eq.(2) where $F_n$ is treated as parameters by considering the screening and correlation effects. $H_{3d-3d}$ includes the Coulomb and exchange interactions in $e_g$ orbitals and the Hund-coupling between $e_g$ and $t_{2g}$ spins [4]. The inter-atomic Coulomb interaction $V(2p_{\gamma_{\theta+}}, 4p_\gamma)$ is not included in the model, in accordance with the previous discussion. Being based on the Hamiltonian, the imaginary part of the scattering factor $((\Delta f_i'')_{xx(zz)} m/\pi |A|^2) = \sum_l |\langle l| P^\dagger_{ix(z)} s_i |0\rangle|^2 \delta(\omega - \varepsilon_l + \varepsilon_0)$ is calculated by the configuration interaction method. In order to check the adequacy of the several parameters, we also calculate the photoemission, inverse photoemission and the optical spectra in the Hamiltonian.

The calculated $(\Delta f_i'')_{\alpha\alpha}$ near the K edge is shown in Fig. 1 (a), where $3d_{3z^2-r^2}$ orbital is occupied. It is noted that the edge of the lowest main peak corresponds to the Mn K edge. The detail structure away from the edge may become broad and be smeared out in the experiments by orverlapping with other peaks which are not included in the calculation. In the figure, the clear anisotropy is shown near the edge where the scattering intensity is governed by $(\Delta f_i'')_{xx}$. Owing to the core hole potential, the main and satellite peaks are attributed to the $|\underline{1s}\ 3d^1_{\gamma_{\theta+}} 3d^1_{\gamma_{\theta-}} 4p^1_{x(z)}\ \underline{2p_{\gamma_{\theta-}}}\rangle$ and $|\underline{1s}\ 3d^1_{\gamma_{\theta+}} 4p^1_{x(z)}\rangle$ states, respectively, although the states are strongly mix with each other. Therefore, the anisotropy in the main peak is caused by $V(3d_{\gamma_{\theta+}}, 4p_\gamma)$ through the Mn $3d$-O $2p$ hybridization. As a comparison, the results in the case where $d_{x^2-y^2}$ orbital is occupied is shown in Fig. 1(b). In the figure, the anisotropy near the edge is entirely opposite to that in Fig. 1(a), i.e., the scattering factor near the edge is governed by $(\Delta f_i'')_{zz}$, owing to the positive value of $V(3d_{x^2-y^2}, 4p_x) - V(3d_{x^2-y^2}, 4p_z)$. In the general case where the occupied orbital is $d_{\gamma_{\theta+}}$, the amplitude and energy position of the main peak in $(\Delta f_i'')_{xx(zz)}$ are shown as functions of $\theta$ in Fig. 2. At $\theta = 2\pi/3$ ($3d_{3y^2-r^2}$) and $5\pi/3$ ($3d_{z^2-x^2}$), $(\Delta f_i'')_{xx}$ and $(\Delta f_i'')_{zz}$ are identical, as expected. Differences of the energy position and the intensity between $(\Delta f_i'')_{xx}$ and $(\Delta f_i'')_{zz}$ are the largest, not at $\theta = 0$ ($3d_{3z^2-r^2}$), but around $\theta = \pi/6$ and $7\pi/6$. It is noted that the anisotropy at $\theta = 0$ ($d_{3z^2-r^2}$) is almost the same as that at $\theta = \pi/3$ ($d_{y^2-z^2}$), although the peak is slightly pushed up in the higher energy region in the former case. This is due to the fact that $V(3d_{3z^2-r^2}, 4p_z) - V(3d_{3z^2-r^2}, 4p_x) = V(3d_{y^2-z^2}, 4p_z) - V(3d_{y^2-z^2}, 4p_x) = 6F_2$.

Murakami et al. analyzed their experimental results in La$_{0.5}$Sr$_{1.5}$MnO$_4$ phenomenologically [7]. In the $(d(x,y,z)/d(y,x,z))$-type orbital ordering which they assumed, the intensity of the forbidden reflection is proportional to the difference of the scattering factor $|(\Delta f)_{xx} - (\Delta f)_{yy}|^2$. It is concluded in the present study that the observed sharp intensity corresponds to the anisotropy of the scattering factor at the edge shown in Fig. 1(a). The anisotropy results from the orbital dependence of the Coulomb interaction. However, even by the present microscopic calculation, it is difficult to determine which orbital ordering, $(d_{3x^2-r^2}/d_{3y^2-r^2})$ or $(d_{z^2-x^2}/d_{y^2-z^2})$, exists, as mentioned above. It is necessary to perform the experiments which determine the tensor elements $((\Delta f_i)_{\alpha\alpha})$ of the structure factors where $\alpha$ is perpendicular to the MnO$_2$ plane in La$_{0.5}$Sr$_{1.5}$MnO$_4$.

The anomalous X-ray scattering may also detect the orbital excitations in the orbital ordered states, that is, the orbital waves [4]. The mechanism of the observation is analogous to the Raman scattering to detect the two orbital waves [25]. Let us start from the antiferro-type orbital ordered state described by $|3d^1_{\gamma_{\theta+}}\rangle_i |3d^1_{\gamma_{\theta-}}\rangle_j$ where $i$ and $j$ denote the nearest neighboring (NN) Mn sites. As mentioned above, the intermediate state in the anomalous X-ray scattering is mainly dominated by the state where a hole is located in the oxygen site between two Mn sites. This state mixes with $|\underline{1s}\ 3d^1_{\gamma_{\theta+}} 3d^1_{\gamma_{\theta-}} 4p^1_{x(z)}\rangle_i$ through the electron transfer between $j$ and the oxygen sites. After the $4p$ electron fills up the $1s$ hole, one $3d$ electron at $i$ site comes back to $j$ site. As a result, the system is relaxed to the final state represented as $|3d^1_{\gamma_{\theta-}}\rangle_i |3d^1_{\gamma_{\theta+}}\rangle_j$, where the orbital state at $i$ and $j$ sites are exchanged, that is, two orbital waves are emitted by the X-ray. Since the E1 transition in the present case is local, the intermediate state accompanied with the charge transfer from O $2p$ orbital to Mn $3d$ orbital promotes the orbital exchange process. In this case, photons with the polarization $x$ bring about the exchange process between $\vec{i}$ and $\vec{i} \pm \hat{y}(\hat{z})$ sites, where $\hat{y}$ is the unit vector in the $y$ direction, as well as $\vec{i}$ and $\vec{i} \pm \hat{x}$ sites. This is in contrast to the conventional Raman process, where photons with polarization $\alpha$ only exchange the orbital states between $\vec{i}$ and $\vec{i} \pm \hat{\alpha}$ sites. Because the spin states in NN sites are also exchanged in this process, in the $(d_{3x^2-r^2}/d_{3y^2-r^2})$ orderings with the A(layer)-type antiferromagnetic structure realized in LaMnO$_3$, the orbital and spin couple and contribute to the inelastic X-ray scattering. The orbital wave in LaMnO$_3$, which has an excitation gap in same order of the band width, is expected to be observed in the region of the order of 100meV [4].

Furthermore, when there is low-lying fluctuations in the orbital degrees of freedom, it is possible to detect by the diffuse scattering in the anomalous X-ray scattering. It is expected that near the orbital ordering temperature in LaMnO$_3$ and La$_{0.5}$Sr$_{1.5}$MnO$_4$ the diffuse scattering owing to the $(d_{3x^2-r^2}/d_{3y^2-r^2})$-type orbital fluctuation becomes remarkable as the critical scattering around $(3/4,3/4,0)$ and $(3,0,0)$, respectively. Another candidate is the orbital fluctuation with $d_{x^2-y^2}$, $d_{y^2-z^2}$ and $d_{z^2-x^2}$ characters in the ferromagnetic metallic phase in La$_{1-x}$Sr$_x$MnO$_3$. The flat dispersions of the orbital fluctuation along the $\Gamma - X$ and other equivalent directions in the Brillouin zone is predicted in this phase [5]. These characteristic dispersions imply the two dimensional character of the orbital fluctuation and will provide the strong anisotropic shape of the diffuse scattering around the fundamental reflection points.

In summary, we have derived the theory of the anomalous X-ray scattering in perovskite manganites and applied to the recent experiments. The dipole transition from Mn $1s$ level to Mn $4p$ level causes the anomalous scattering. The Hartree potential without cubic symmetry due to $e_g$ electron is identified as the main mechanism of the anisotropy of the scattering factor. The anomalous X-ray scattering is promising as a probe to study the orbital degrees of freedom in not only manganites but other TMO.

The authors would like to thank Y.Endoh for suggesting this study and encouraging at every stage of the work, Y.Murakami for providing the experimental data prior to publication, and W.Koshibae for valuable discussions. This work was supported by Priority Areas Grants from the Ministry of Education, Science and Culture of Japan, and CREST ( Core Research for Evolutional Science and Technology Corporation) Japan.

Figure captions.

Fig. 1: The imaginary part of the scattering factor $((\Delta f_i'')_{xx(zz)} m/\pi |A|^2)$ in the case where the following orbital is occupied: (a) $\theta = 0$ ($d_{3z^2-r^2}$), and (b) $\theta = \pi$ ($d_{x^2-y^2}$). The straight and broken lines show $(\Delta f_i'')_{xx}$ and $(\Delta f_i'')_{zz}$, respectively. The parameter value of $\delta$ in Eq.(1) is chosen to be $0.5eV$. The origin of the energy is taken to be arbitrary.

Fig. 2: The amplitudes and energy position of the main peak in the scattering factor $((\Delta f_i'')_{xx(zz)} m/\pi |A|^2)$ as functions of $\theta$. The straight and broken lines show $(\Delta f_i'')_{xx}$ and $(\Delta f_i'')_{zz}$, respectively. The origin of the energy is taken to be arbitrary.